\documentclass[showpacs,floatfix,eqsecnum,preprint,nofootinbib]{revtex4}

\usepackage{graphics}

\usepackage{bm}

\let\d=\delta
\let\D=\Delta

\let\z=\zeta

\let\r=\rho

\let\F=\Phi

\let\w=\omega
\let\o=\omega
\let\om=\omega

\begin{document}
\title{Newton's second law versus modified-inertia MOND: a test using the high-latitude effect}
\author{A.~Yu.~Ignatiev}
\email{a.ignatiev@ritp.org}
 \affiliation {{\em Theoretical Physics Research Institute, Melbourne 3163,} \\ {\em   Australia.}}
\begin{abstract}
The modified-inertia MOND is an approach that proposes a change in Newton's second law at small accelerations as an alternative to dark matter.
Recently it was suggested that this approach
can be tested in terrestrial laboratory experiments. One way of doing the test is based on the Static High-Latitude Equinox Modified inertia (SHLEM) effect:
around each equinox date,
2 spots emerge on the Earth where static bodies experience spontaneous displacement due to the  violation of  Newton's second law required by the modified-inertia MOND.
Here, a detailed theory of this effect is developed and estimates of the magnitude of the signal due to the effect are obtained. 
The expected displacement of a mirror in a gravitational wave  interferometer is found to be about $10^{-14}$ m. Some experimental aspects  of the proposal
are discussed.

\end{abstract}
\maketitle
\section{Introduction}
In this paper I examine the experimentally testable consequences of the hypothesis  that  Newton's second law should be changed for small accelerations. This hypothesis - called the modified  Newtonian dynamics (MOND) with modified inertia - has been proposed by Milgrom \cite{m0} as an alternative to the more conventional Dark Matter explanation of the shape of the galactic rotation curves.

The main assumption of the modified-inertia MOND
is that Newton's second  law should be modified to read
 \begin{equation}
\label{}
\mathbf{F}=m\mathbf{a}\mu(a/a_0),
\end{equation}
where $a_0$ is a fundamental acceleration of the order of $10^{-10}\; m\,s^{-2}$ and $\mu$ is a function satisfying the two conditions: $\mu(a/a_0) \rightarrow 1$ at $a\gg a_0$ and\footnote{The difference between this 
asymptotic condition and that of Ref. \cite{prl} is due to a misprint in the latter.} $\mu(a/a_0) \rightarrow a/a_0$ at $a\ll a_0$.  (More details about $\mu$ and $a_0$ are in Sec. V and VI.)

Further, the universal gravitation law is assumed to keep its conventional form
\begin{equation}
\label{}
F=\frac{Gm_1m_2}{r^2}.
\end{equation}

At this stage, the modified-inertia MOND
 is only formulated in the context of Newtonian physics, i.e., a flat-space picture, which will be assumed throughout this paper.

At the moment the MOND approach is attracting constantly growing interest. Various  astrophysical aspects of this approach  are under active discussion (see \cite{w} and the last Ref. \cite{m0} for a comprehensive bibliography).  However, until lately the analysis of terrestrial (as opposed to astrophysical) consequences of MOND has been missing in the literature. 
The unusual smallness of the acceleration $a_0$ makes it very difficult to think of a laboratory test. It also explains why we do not see any deviations from Newton's second law under ordinary circumstances.
Surprising first results of such analysis have recently been  described
in \cite{prl}.

One of these results is the existence of a new effect that  has been called Static High-Latitude Equinox Modified inertia or SHLEM: around each equinox date,
2 spots emerge on the Earth where static bodies experience spontaneous displacement due to the  violation of  Newton's second law required by the modified-inertia MOND. The laboratory observation of this effect may be accessible to current experimental capabilities. 

Consequently, it could serve as a basis for  a  proposal to test the validity of Newton's second law for small accelerations in a laboratory based experiment---the first of its kind. In fact, such a test could become a crucial experiment in view of the lack of conclusive astrophysical evidence either in favour or against the  modified Newtonian dynamics.

Here, a detailed theory of this effect is developed and estimates of the magnitude of the signal due to the effect are obtained. 
The expected displacement of a mirror in a gravitational wave  interferometer is found to be about $10^{-14}$ m.

The nature of inertia has long been one of the most fundamental puzzles of physics. In recent times, this puzzle has widened to include also the problem of  the origin of mass  in the Standard Model. Despite concerted theoretical efforts over decades, the path to the solution is still to be found.

Thus the pursuit of MOND as an alternative explanation of the `dark matter' puzzle seems well motivated. If true, it will bring about the need to revise foundations of modern physics.

The paper is organized as follows.
To make the exposition more understandable and self-contained,
Sections II and III review and extend the background material from \cite{prl}.
Section IV  describes the SHLEM effect qualitatively.
In Sec. V and VI the equation of motion is derived and its solution is obtained. In Sec. VII some experimental suggestions are discussed.
Finally, the  conclusions are presented in Sec. VIII.

\section{Acceleration: `absolute' or  `relative'?}

First I  emphasize that to obtain laboratory-testable predictions,
  MOND needs to be formulated not only in inertial reference systems, but also in non-inertial systems as well \cite{prl}. (In the MOND context all laboratory reference systems should be considered as  non-inertial.) Because  the dynamical law is modified depending on the acceleration, the transition between inertial and non-inertial systems in MOND becomes less straightforward than in the conventional mechanics. 

Of particular interest are  transformation properties
of  $a_0$.  
 Logically, at least 2 options could be imagined. First, one can assume that the fundamental acceleration that determines the onset of the MOND regime equals $a_0$ only in the inertial reference systems.
Second, it could be assumed that $a_0$
 is invariant under transformations from inertial to non-inertial systems. One would expect that these 2 types of theories would lead to drastically different experimental predictions.

For instance, the first  type of theory requires that the MOND regime is reached as soon as the test body moves with a tiny acceleration $\alt a_0$ \emph{with respect to the Galactic reference frame}\footnote{Although the choice of the Galactic reference frame
 is a natural one, 
 a question can be raised whether alternative choices
 are possible.
 For instance, what if we take a reference frame with the origin at the center of mass of the Local Group of galaxies? Fortunately, that would not   significantly affect the results because the acceleration due to the neighbouring galaxies is much smaller than $a_0$. For example, the acceleration due to the Andromeda galaxy is less than $10^{-12}\; m\, s^{-2}$.}.

On the other hand,  the second type  of theory implies that in order to reach the MOND regime, we should try to ensure that the test body moves with a tiny acceleration $\alt a_0$ \emph{with respect to the laboratory reference frame}. 

However, it  has been pointed out  that
the second version (invariant acceleration $a_0$) is not self-consistent \cite{prl}. The reason is that the invariance of $a_0$ is inconsistent with the kinematical rules of acceleration addition. Indeed, let us take two reference frames: an inertial S, and a non-inertial S$'$ that is in translational motion
with acceleration ${\bf b}$ relative to S.  Then, in the system S the equation of motion will be 
\begin{equation}
\label{a1}
\mathbf{F}=m\mathbf{a}\mu(a/a_0),
\end{equation}
where $\bf{a}$ is the acceleration of a test body in the system S.  If we assume that $a_0$ is invariant, then the equation of motion in the system S$' $ is
\begin{equation}
\label{a2}
\mathbf{F}=m\mathbf{a'}\mu(a'/a_0)+m{\bf b},
\end{equation}
where ${\bf a'}={\bf a}-{\bf b}$
is the acceleration of the test body in the system S$'$. However, equations (\ref{a1}) and (\ref{a2}) cannot hold simultaneously for all $\bf{a}$ and $\bf{b}$: for instance, if we put ${\bf a}=0$ then we obtain
\begin{equation}
\label{}
m\mu(b/a_0)=m
\end{equation}
for all $\bf{b}$ which means that $\mu(z)=1$ for all $z$. Thus the invariant-$a_0$ version of MOND is inconsistent.

In addition, this version has been ruled out experimentally \cite{f}.
In what follows, only the first version will be considered.

\section{Condition of entry into SHLEM regime}

We will now recall what conditions must be realized in order to obtain  the SHLEM effect for  test bodies at rest in the ground-based laboratory \cite{prl}.

This question is easy to answer in the inertial system $S_0$. (It is the system with the origin in the centre-of-mass of our Galaxy and the axes pointing to certain far-away quasars). In this system, we should ensure that the test body moves with a tiny acceleration ${\bf a}_{gal}$ with respect to $S_0$:
\begin{equation}
\label{0}
\mathbf{a}_{gal}\approx 0.
\end{equation}
In this Section, the $\approx$ sign will mean that the difference between the left-hand side and the right-hand side of an equation is much less than the characteristic MOND acceleration $a_0$.

Next, we are going to the laboratory system with the help of 
\begin{equation}
\label{0a}
\mathbf{a}_{gal}\approx\mathbf{a}_1(t)+{\bm\om}\times(\bm{\om}\times(\mathbf{r}+\mathbf{r}_1))+\mathbf{a}_2,
\end{equation}
where $\mathbf{a}_1$ is the acceleration of the Earth's centre with respect to the heliocentric reference frame,
$\bm{\om}$ is the angular velocity of the Earth's rotation, $\mathbf{a}_2$ is the Sun's acceleration with respect to $S_0$;
$\mathbf{r}$ is the position of the test body with respect to the laboratory reference frame;  $\mathbf{r}_1$ is the position vector of the origin of the lab frame with respect to the terrestrial frame with the origin at the Earth's centre\footnote{As practical, high-precision realizations of these intermediate frames one can take the International Celestial Reference System (ICRS) \cite{icrs} and the International Terrestrial Reference System (ITRS) \cite{itrs}.}. A number of terms have not been written out in Eq.~(\ref{0a}) on account of their smallness. They include terms due to: the Coriolis acceleration of the Sun, the length-of-day (LOD) variation, precession and nutation of the Earth's rotation axis, polar motion and Chandler's wobble (see Table 1 for their approximate magnitudes).
\begin{table}[htdp]
\caption{Accelerations that can be ignored.}
\begin{ruledtabular}
\begin{tabular}{cc}
Source& Approximate magnitude, $\rm m/s^2$\\ \hline
``Galactic Coriolis''&$ 2\times 10^{-11}$\\
short period LOD variation&$3\times 10^{-12}$\\
annual LOD variation&$ 1\times 10^{-13}$\\
precession&$ 2\times 10^{-14}$\\
secular increase of LOD&$3\times 10^{-15}$\\
nutation (main term, epoch 1900,0)&$9\times 10^{-20}$\\
 Chandler's wobble&$1\times 10^{-20}$\\
 annual pole motion&$4\times 10^{-21}$\\
  secular polar motion&$3\times 10^{-24}$\\
\end{tabular}
\end{ruledtabular}
\label{default}
\end{table}

From Eq.~(\ref{0}) and Eq.~(\ref{0a}) we obtain the necessary and sufficient condition for realisation of the MOND regime in the laboratory:

\begin{equation}
\label{2.0}
 \mathbf{a}_s(t)+\bm{\om}\times(\bm{\om}\times \mathbf{r}_1)\approx 0,
\end{equation}
where I  have introduced $ \mathbf{a}_s =  \mathbf{a}_1+  \mathbf{a}_2$ for convenience and put $\mathbf{r}=0$ (without loss of generality).

We note that this equation has no solutions unless $\mathbf{a}_s$ is orthogonal to $\bm{\om}$, so we must first look for those instants $t_p$ when
\begin{equation}
\label{55}
a_{s\parallel}(t_p) \approx 0 \;  or \;  a_{s\parallel}(t_p)| \ll a_0,
\end{equation}
where $a_{s\parallel}=(\mathbf{a}_s\bm{\om})/\om$.
A continuity argument shows that this equation has at least 2 solutions during each year. Indeed, at the instant of a (nothern) summer solstice $a_{s\parallel} > 0$ whereas at the instant of  a  winter solstice $a_{s\parallel} < 0$. Therefore, there must be at least one instant during autumn and one instant during spring when $a_{s\parallel} = 0$ {\em exactly}. Neglecting the effects due to the Moon and planets, these instants would coincide exactly with the autumnal and vernal equinoxes. In reality, the instants will be shifted from the equinoxes. However, the above `existence theorem' guarantees that these instants $t_p$ can be found with astronomical precision through a straightforward but time-consuming procedure using the lunar and planetary ephemerides. In addition, one can show that the off-equinox shift, in any case, should be less than a few days.

Once  $t_p$ is found and plugged into Eq.~(\ref{2.0}), the corresponding solution for the laboratory location is
\begin{equation}
\label{key}
\mathbf{r}_{1\perp}= \frac{\mathbf{a}_s(t_p)}{\om^2}.
\end{equation}

This  key relation allows us to find both the latitude and the longitude of the right spot. If we again ignore the lunar and planetary effects, the relevant magnitude is $|\mathbf{a}_s(t_p)| \simeq 0.00593  \; m\,s^{-2}$ which gives the required latitude  $\phi\simeq\pm 79^o50'$. As for the longitude, it would generally vary from year to year. For instance, on the autumnal equinox of September 22, 2008 these spots would be at $56^o$ W---one in Greenland, ($79^o50'$ N), another in Antarctica ($79^o50'$ S). Predictions for the years 2008-2019 are shown in Table 2. The dates and times of the equinoxes are taken from Ref. \cite{usno}.
\begin{table}[htdp]
\caption{Longitudes of the `SHLEM spots' calculated in the `Earth-Sun' approximation.}
\label{}
\begin{ruledtabular}
\begin{tabular}{ccc}
Date & Time (UT)& Longitude\\ \hline
2008 Mar 20	&	5:48	&	$	93^{\circ}	\,	\rm	E	$	\\
2008 Sept 22	&	15:44	&	$	56^{\circ}	\,	\rm	W	$	\\
2009 Mar 20	&	11:44	&	$	4^{\circ}	\,	\rm	E	$	\\
2009 Sept 22	&	21:18	&	$	139^{\circ}30'	\,	\rm	W	$	\\
2010 Mar 20	&	17:32	&	$	83^{\circ}	\,	\rm	W	$	\\
2010 Sept 23	&	3:09	&	$	132^{\circ}45'	\,	\rm	E	$	\\
2011 Mar 20	&	23:21	&	$	170^{\circ}15'	\,	\rm	W	$	\\
2011 Sept 23	&	9:04	&	$	44^{\circ}	\,	\rm	E	$	\\
2012 Mar 20	&	5:14	&	$	101^{\circ}30'	\,	\rm	E	$	\\
2012 Sept 22	&	14:49	&	$	42^{\circ}15'	\,	\rm	W	$	\\
2013 Mar 20	&	11:02	&	$	14^{\circ}30'	\,	\rm	E	$	\\
2013 Sept 22	&	20:04	&	$	121^{\circ}	\,	\rm	W	$	\\
2014 Mar 20	&	16:57	&	$	74^{\circ}15'	\,	\rm	W	$	\\
2014 Sept 23	&	2:29	&	$	142^{\circ}45'	\,	\rm	E	$	\\
2015 Mar 20	&	22:45	&	$	161^{\circ}15'	\,	\rm	W	$	\\
2015 Sept 23	&	8:20	&	$	55^{\circ}	\,	\rm	E	$	\\
2016 Mar 20	&	4:30	&	$	112^{\circ}30'	\,	\rm	E	$	\\
2016 Sept 22	&	14:21	&	$	35^{\circ}15'	\,	\rm	W	$	\\
2017 Mar 20	&	10:28	&	$	23^{\circ}	\,	\rm	E	$	\\
2017 Sept 22	&	20:02	&	$	120^{\circ}30'	\,	\rm	W	$	\\
2018 Mar 20	&	16:15	&	$	63^{\circ}45'	\,	\rm	W	$	\\
2018 Sept 23	&	1:54	&	$	151^{\circ}30'	\,	\rm	E	$	\\
2019 Mar 20	&	21:58	&	$	149^{\circ}30'	\,	\rm	W	$	\\
2019 Sept 22	&	13:30	&	$	22^{\circ}30'	\,	\rm	W	$	
\end{tabular}
\end{ruledtabular}
\end{table}

 The account of lunar perturbation can significantly change the longitude, but the latitude prediction is much more robust: it would not change by more than $\sim 6'$, or $10\; km$.

\section{The general picture of motion}
The signature of the SHLEM effect would be a spontaneous displacement of the test body occurring 
around the instant $t_p$ defined by Eq.~(\ref{55}), and we are now ready to start calculating its magnitude. 
The qualitative scenario runs as follows: around the instant $t_p$ the test body and the reference body
move according to two different laws of motion. Roughly speaking, MOND  makes the test body to `lose' its mass while the the reference body
keeps its normal mass (because it lies outside of the MOND regime). Therefore the test body will shift with respect to  the reference body
 by a tiny but non-zero, time-dependent distance $x(t)$.

Throughout the paper I will  
work in the `Earth-Sun' approximation in which the corrections due to the Moon and planets 
are ignored. A question arises: how would account of these corrections change the results? As was shown previously \cite{prl}, the time and location of the SHLEM event can change slightly, but the effect itself will survive because its existence is based on topological arguments.

Another worry is whether we can use Newtonian, not modified inertia mechanics to calculate the instant $t_p$ \cite{n}. To address this, we note that this calculation rests on the fact that the motion of the Earth as a whole obeys the Newtonian mechanics. Of course, within the MOND approach this is not absolutely true. Moreover, the approach developed in \cite{prl} and here allows one to calculate the MOND corrections to the value of $t_p$ as precisely as one wishes. However, these corrections appear to be so tiny that they can be completely neglected even without their precise calculation. Indeed, they would, in any case, include the suppressing factor of 
$M_{MOND}/M_{Earth}$ where $M_{MOND}$ is the mass of the part of the Earth affected by MOND. 

The MOND-affected mass can be (generously) bounded from above as follows
\begin{equation}
\label{}
M_{MOND}\alt 2R_E \sin{\phi} (v_r\d t)^2 \r_E,
\end{equation}
where $R_E$, $\r_E$, and $v_r$ are the Earth radius, density, and linear rotation speed at the latitude $\phi\approx 80^o$. The time scale $\d t$ was defined in \cite{prl} as

\begin{equation}
\label{}
\d t\sim (a_0/a_s)(4\epsilon/T)^{-1}\sim1\; s
\end{equation}
where $\epsilon=23^o27'=0.41$, $T=1\; yr$.

Altogether, the suppressing factor is
\begin{equation}
\label{}
M_{MOND}/M_{Earth}\alt 10^{-10},
\end{equation}
which means that for the purposes of calculating the SHLEM instant  and coordinates of the SHLEM spot, the Newtonian mechanics can be used quite safely.

\section{Equation of motion}
We start by working in an inertial reference frame $S_0$ first. 
Denote by ${\bf f}$ the total physical force acting on the unit mass.   Then the MOND equation of motion is
\begin{equation}
\label{18}
{\bf f}=\ddot{\bf r}\mu(|\ddot{\bf r}|/a_0).
\end{equation}

Although the present treatment can be used with any interpolating function $\mu$,  to obtain a definite
result one needs to fix its concrete form; 
 the standard choice has been \cite{m0,ma,ba}
\begin{equation}
\label{18a}
\mu(z)=\frac{z}{\sqrt{1+z^2}}.
\end{equation}
 It is a matter of current debate \cite{ser,ior} whether this function satisfies the constraints derived from the precision solar system data \cite{pit}.
In any case, a new function can only be introduced after a reanalysis of the galactic rotation curves and a corresponding reestimate of the acceleration scale $a_0$. (For instance, it would be inconsistent to use a new interpolating function with the old value of $a_0$.) For this reason, at the moment
we have no choice for $\mu$ other than (\ref{18a}), and 
 it will be used from now on.

Solving Eq. (\ref{18})  for $\ddot{\bf r}$ we obtain
\begin{equation}
\label{20}
\ddot{\bf r}={\bf f}\Phi({\bf f}^2/a_0^2),
\end{equation}
where
\begin{equation}
\label{}
\Phi(\z)=\sqrt{  \frac  {1+\sqrt{1+\frac{4}{\z}} } {2}  }.
\end{equation}

Next, we go into the laboratory reference frame by adding the non-inertial accelerations ${\bf f}_{in}$:
\begin{equation}
\label{}
\ddot{\bf r}={\bf f}\Phi({\bf f}^2/a_0^2)+{\bf f}_{in}.
\end{equation}

To obtain a more specific equation, let us make simplifying assumptions similar to those usually made when analyzing the response of the gravitational wave detectors (these assumptions  will be justified in the next Section when the solution of the equation of motion is obtained):

---the body interacts with its environment (such as a support, suspension etc.) through an elastic force with negligible dissipation:
\begin{equation}
\label{}
{\bf f}_{el}=-\omega_0^2{\bf r}.
\end{equation}

---random forces (e.g., due to thermal and vibrational noises) are ignored.\footnote{The account of these would require a detailed knowledge of the experimental setup. If a gravitational-wave type of detector is considered, the account can be made using methods developed in that area.}

---the test body is treated as a point mass, i.e., we ignore a small variation of $f$ over its volume.

---terms of the second and higher order in the small parameter $\w (t-t_p)$ are neglected.

---displacement of the reference body  is negligible. (The reference body is the body that plays the role of the origin of the laboratory reference frame. For example, in interferometers the reference body is one of the mirrors.)

Therefore, the total physical force will consist of the sum of gravitational forces and the elastic coupling:
\begin{equation}
\label{}
{\bf f}=\sum{{\bf f}_{gr}}+{\bf f}_{el}={\bf f}_{\Sigma gr}-\omega_0^2{\bf r},
\end{equation}
and the resulting equation of motion will take the form:
\begin{equation}
\label{}
\ddot{\bf r}+\omega_0^2{\bf r}\Phi={\bf f}_{\Sigma gr}\Phi +{\bf f}_{in}.
\end{equation}
Let us now introduce the following coordinate system within the laboratory   reference frame: $x$ is the West-to-East axis, $y$ ---the South-to-North axis, and $z$ is the vertical axis. Then in our approximation the projection of the above equation on these axes 
will give
\begin{equation}
\label{main}
\ddot{x}+\omega_0^2{x}\Phi\simeq f(t)\Phi-f(t),\;\; \ddot{y}\simeq 0,\;\;\ddot{z}\simeq 0,
\end{equation}
where
\begin{equation}
\label{}
f(t)\simeq a_s\o(t-t_p),\;\; \F\simeq \F[(f-\omega_0^2x)^2/a_0^2].
\end{equation}
Thus we can ignore motion along $y$ and $z$ axes and our problem becomes one-dimensional.

\section{Solution of the equation of motion}

The obtained non-linear equation of motion is best solved numerically.
But first we need to fix the frequency $\o_0=2\pi f_0$. It is determined by the suspension design: for example, $f_0=0.65$ Hz in the case of LIGO\footnote{VIRGO has a very close frequency: $f_0=0.60$ Hz.}. Therefore,  $\o_0=4.1\;\rm s^{-1}$  will be used.

At present there is some uncertainty in the magnitude of $a_0$, the fundamental parameter of the theory. For instance, in Refs. \cite{ma} and \cite{ba} it was found that $a_0=(1.5\pm0.7)\times 10^{-10}\;m\,s^{-2}$ and $a_0=(1.35\pm0.51)\times 10^{-10}\;m\,s^{-2}$, respectively.\footnote{Ref. \cite{ba} also quotes a slightly different value: $a_0=(1.21\pm0.27)\times 10^{-10}\;m\,s^{-2}$. It is obtained if one outlier galaxy is removed from the sample on the basis of a possible error in the determination of its distance. The value $a_0=1.2\times 10^{-10}\;m\,s^{-2}$ is generally assumed in the literature.}
Because of that, 
we will first assume that $a_0=2\times 10^{-10}\;m\,s^{-2}$
 and then repeat the calculation with $a_0=1.2\times 10^{-10}\;m\,s^{-2}$ and $a_0=1\times 10^{-10}\;m\,s^{-2}$.

\begin{figure}[htbp]
\begin{center}
\includegraphics{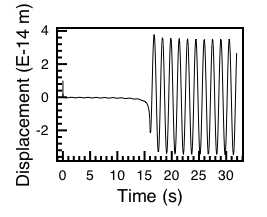}
\caption{Displacement of the mirror as function of time. The instant $t_p$ corresponds to $t=16$ s.}
\label{sol}
\end{center}
\end{figure}

The resulting solution for $a_0=2\times 10^{-10}\;m\,s^{-2}$ is shown in Fig.~\ref{sol}. 
The displacement amplitude has a maximum of $\simeq 3.8 \times 10^{-14}$ m. This maximum is reached $\simeq 0.76$ s 
after the SHLEM instant $t_p$.

Having obtained the solution we can now fully justify the assumptions made earlier while deriving the equation of motion. First, the damping  can indeed be completely neglected given that the relevant quality factor is $Q=\rm few\times 10^3$ \cite{lo}.

Secondly, the linear time dependence in $f(t)$ can only be used for  `short' times $t$ so that $\o (t-t_p)\ll 1$, where $\o$ is the Earth's angular velocity. In other words, the time intervals must be much less than $24\;\rm h/2\pi\simeq 4$ hours. For our solution this condition obviously holds by a wide 
margin.

Thirdly, the neglect of transverse motion is justified by the same condition $\o (t-t_p)\ll 1$.

Fourthly, let us confirm that the size of the mirror  does not matter.
Indeed, the difference in centrifugal accelerations that act on the mirror edges is
\begin{equation}
\label{}
\D a=a_s\frac{x}{R_E \cos{\phi}},
\end{equation}
where $x$ is the distance between edges, $R_E$ is the Earth's radius and $\phi$ --- the lab latitude. In the case of LIGO, depending on the mirror orientation, $x$ can take the values of 9.3 cm (the mirror thickness) or  25 cm (the mirror diameter). To ensure that this difference is insignificant we must require that it be less than the `driving force' $f(t)=a_s\o(t-t_p)$ in Eq.~(\ref{main}). It is easy to check that this requirement is always fulfilled except during a very short interval 
\begin{equation}
\label{}
\D t \sim \frac{x}{R_E \o\cos{\phi}}\sim 3\times 10^{-3}\left(\frac{x}{25\; cm}\right) s
\end{equation}
around $t_p$. But the contribution of this interval to the total displacement of the mirror  is quite negligible as can be seen from Fig. 1. This justifies our treatment of the mirror as a mass point.

Finally, we can check if the displacement of the second (reference) mirror can be neglected. At the SHLEM latitude $\phi \approx 80^o$ the Earth rotates with a linear speed of $v_r=R_E \o\cos{\phi}\approx 80$ m/s (eastward). This means that
the SHLEM spot runs on the Earth's surface with the same speed (westward). Thus, if a reference mirror is farther than 80 m from the test mirror, then the reference mirror's motion can be neglected during an observation interval of about 1 s around the SHLEM instant $t_p$. Otherwise (i.e., if the distance is shorter than 80 m or the observation interval is longer than 1 s) the motion of the reference mirror must be taken into account. In that case, the signal will be equal to the difference between the two mirrors' displacements.

Now, it is of interest to see how sensitive is the obtained solution to the variation of $a_0$. If we use the value $a_0=1\times10^{-10}\;m\,s^{-2}$ instead of $a_0=2\times10^{-10}\;m\,s^{-2}$ then the maximum of the displacement amplitude will drop  to  
$\simeq 0.95 \times 10^{-14}$ m
(i.e., by a factor of 4). Finally, if the currently favored value $a_0=1.2\times10^{-10}\;m\,s^{-2}$ is used, the maximum will be at $\simeq 1.4 \times 10^{-14}$ m. In all cases considered, it takes approximately the same time ($\simeq 0.76\;s$ counting from the SHLEM instant $t_p$) to reach the maximum.

\section{Experimental  considerations}

A variety of approaches to the experimental searches for the modified inertia effects have been proposed in Ref.~\cite{prl}.
Because we want to detect a tiny displacement/acceleration, it is natural to turn to the vast and vigorous area of gravitational experiments.
In particular, it was pointed out
that the existing gravitational wave detectors could be a good starting point in designing the experiment.\footnote{Recently, an analog of the SHLEM effect for the satellites' motion was considered by McCulloch \cite{mc}
starting from a different way of modifying inertia \cite{mc0}.}

At present, there are 2 types of gravitational wave facilities: the interferometers and resonators. The interferometers are generally more sensitive.  So it is reasonable to start our discussion with them\footnote{The resonant detectors can be analysed in a similar way, but technically there are some differences which will not be considered here.}. 

The most sensitive of the currently operating interferometers are LIGO and Virgo \cite{ligo-s,ligo-eps,virgo-s,virgo-eps}.
The heart of such a detector is a suspended 
mirror whose position can be monitored
with an ultrahigh accuracy $\d l \sim 10^{-18}$ m. This is achieved by using a Michelson-type of interferometer with a long arm length: $L=4$ km and $L=2$ km for LIGO, or $L=3$ km for Virgo. The 
detector's sensitivity 
is determined by the dimensionless ratio $\d l/L \sim 10^{-21}$ which can be translated into the  dimensionless amplitude of a detectable gravitational wave.
In some respects, the expected SHLEM signal would be similar to the signal expected from a
gravitational wave. 
However, there is a crucial difference between the two. The SHLEM signal is characterized by {\em dimensional} 
quantity---displacement of the test body $\d l$---rather than a dimensionless ratio $\d l/L$.

As a consequence, {\em having long arm length $L$ is  unnecessary for us} as it does not affect the detector sensitivity  to the SHLEM signal.
This opens up an interesting opportunity of considering \it a 
`short-arm LIGO/Virgo-like' detector \rm
that would be based on the same ideas and technical know-how as LIGO/Virgo themselves, but would have a much smaller size that would allow it to be more easily transported and installed in a required location.

In other words, we can imagine a setup that would be similar
to LIGO/Virgo, but with a much shorter arm length
 (as is the case for TAMA300 \cite{tama},  GEO \cite{geo}, or ACIGA/AIGO \cite{aigo}, where the respective arm lengths are  about 300 m, 600 m, and 80 m).
 LIGO's  Prototype Interferometer on the Caltech campus has even a shorter arm length of 40 metres.\footnote{A lower bound on the arm length would be set by the desirability to keep the second mirror unaffected by SHLEM. As discussed at the end of Sec. VI, one way to do this is to make the distance between mirrors large enough. It is hard to work out the precise lower bound in advance because it would depend on the orientation of the arm and other unknowns.}

It is quite possible that additional effort would be required to ensure that the shorter arm-length does not compromize the overall sensitivity of the detector. For example, the issue of high laser power in a short arm would have to be carefully analyzed. At this stage, though, it seems premature to go deeper into these details.

To assess the future potential of an interferometer setup,
we have to take into account the plans \cite{lo} to upgrade both LIGO and Virgo in the near future.
The second generation detectors (Enhanced LIGO and Virgo+) will have their sensitivities increased two- or three-fold by 2009.
At the next stage (Advanced LIGO, Advanced Virgo, 2014) the increase of sensitivity will be ten-fold.
Further, the third generation underground detectors, such as the Einstein gravitational-wave Telescope (E.T.), are currently under active discussion.

Thus it does not seem unrealistic to think
that a suitable
laboratory based on the LIGO/Virgo technology can be conceived
in which the position of the mirror can be monitored with an accuracy sufficient for detecting the SHLEM effect.

In that case, how should one analyze the data? Because of a similarity between a SHLEM signal and a gravitational wave, 
the approach to this problem could be similar
to that used in Refs. \cite{ligo04,ligo05,ligo06,ligo07,virgo06,virgo07,
lv}. In particular, the paper \cite{lv}  
by the joint LIGO-Virgo working group 
 gives a detailed analysis of various statistical procedures aimed at detecting gravitational wave bursts using  LIGO and Virgo facilities. As an input, astrophysical calculations are used which give the waveforms and amplitudes of signals corresponding to different sources. 

Since the distance, position, and nature of the source are not known in advance, one has to take into account many possibilities which lead to considerable uncertainty in the input data. By contrast, in our case, the input is unique: it is as if someone were able to predict everything about the burst--the source, its distance, right ascension and declination, and even the exact time of the burst. It makes the statistical analysis rather more certain and robust compared to the analysis required by a largely uncertain gravitational burst.

The complete information about the profile of the signal in our case is contained in Eq.~(\ref{main}) and Fig.~1 that give the time-dependence of the 
test body's displacement. If one is interested in the signal spectrum, it can 
be obtained from the same equation.

Based on this information, the candidate filters can be 
constructed via methods similar to those employed in Ref. \cite{ligo04,ligo05,ligo06,ligo07,virgo06,virgo07,
lv}. Their performance  can then be tested using simulations with the  signal shape known from Eq.~(\ref{main}) and Fig.~1 and thus 
the optimal filter can be selected.

As is well known, the coincidence of signals in different detectors plays an important role in the strategies of the gravitational wave searches. In our case, various coincidence and anticoincidence schemes can be conceived due to the strict localization of the effect in time and space. For example, one can use the scheme in which the first detector is placed in the SHLEM spot in the northern hemisphere while the second detector is located symmetrically in the southern hemisphere. 

We can therefore 
hope that the chances of detecting the SHLEM signal 
could be better than the chances of detecting a gravitational wave burst  of a similar magnitude.

Because the effect can only be observed at high latitudes, a question can be raised \cite{n} if  icebergs floating in the ocean nearby (say, $D\simeq10$ km away) can trigger  false alarms by creating an excess of gravity of the order of $a_0$.

To assess this possibility, we note that the short time scale  of the effect needs to be taken into account. Namely, to create a false alarm, external source must provide a \em variable \rm gravity: its variation $\d g_{net}$ over the interval $\d t\sim 1$ s must be of the order of $a_0$. 
Here $g_{net}$ is the net excess gravity which the difference between the gravity due to the ice and the `negative' gravity due to the displaced water: 
\begin{equation}
\label{}
g_{net}=g_{ice}-g_{water}\simeq g_{ice}\left(\frac{d}{D}  \right)^2
\simeq g_{ice}\left(\frac{h}{D}  \right)^2,
\end{equation}
where $d$ is the distance between the centers of mass of the ice and the displaced water, $h$ is the hight of the above-water tip of the iceberg.  The exact relation between $d$ and $h$ depends on the iceberg's shape. For instance, a cubic shape  gives $d=h/2$. As we will see shortly, the `shape factor'  would not greatly affect the final result so  $d\sim h$ 
will be assumed.

Next, suppose that the iceberg's velocity is $v$. Then during the time $\d t$ it will shift by the distance  $\d D\sim v\d t$ relative to the lab. That will lead  to gravity variation of the order of 
\begin{equation}
\label{}
|\d g_{net}|\sim \frac{GMh^2}{D^5}\d D\sim  \frac{GMh^2}{D^5} v\d t.
\end{equation}
Requiring $|\d g_{net}|\sim a_0$ leads to the following condition:
\begin{equation}
\label{}
\frac{Mh^2 v}{D^5}\sim \frac{a_0}{G\d t}.
\end{equation}
Therefore, the iceberg mass, tip height, and velocity must satisfy 
\begin{equation}
\label{}
Mh^2v\simeq 3\times 10^{20}\; \rm kg\, m^3\,s^{-1}.
\end{equation}
Such parameters appear to be unrealistically large. Indeed, a very large iceberg would have a mass of about $M\sim 10^{11}$ kg, with $h\sim 100$ m \cite{iip} so the product $Mh^2 v$ falls short by several orders of magnitude. Thus icebergs cannot trigger false alarms.

\section{Conclusions}
Astrophysically inspired laboratory-based experiments such as dark matter searches have become a common part of the physics landscape. The laboratory tests of MOND is a much needed, complementary activity. It is a new virgin territory waiting to be explored.

This paper shows that methods and installations used in the gravitational wave research are likely to be useful also in the new area of searching for modified inertia and the SHLEM effect.

Helpful discussions with L.A.Ignatieva, V.A.Kuzmin, M.E.Shaposhnikov, and H.S.Zhao are gratefully acknowledged.

I am also thankful to E.Adelberger, P.Ball, O. Bertolami, J.Cartwright, D.Castelvecchi, E.Gallo, J.Lucentini, M.Marquit, Z.Merali, and O.M.Moreschi for their interest in the subject matter and stimulating comments.

\end{document}